# Estimate Metabolite Taxonomy and Structure with a Fragment-Centered Database and Fragment Network

Hansen Zhao[1], Xu Zhao[1], Huan Yao[1], Jiaxin Feng[1], Sichun Zhang[1*], Xinrong Zhang[1*]

[1]Department of Chemistry, Tsinghua University, Beijing, China
*Corresponding author: xrzhang@mail.tsinghua.edu.cn; sczhang@mail.tsinghua.edu.cn

**Abstract:** Metabolite structure identification has become the major bottleneck of the mass spectrometry based metabolomics research. Till now, numbers of mass spectra databases and search algorithms have been developed to address this issue. However, two critical problems still exist: the low chemical component record coverage in databases and significant MS/MS spectra variations related to experiment equipment and parameter settings. In this work, we considered the molecule fragment as basic building blocks of the metabolic components which had relatively consistent signatures in MS/MS spectra. And from bottom-up point of view, we built a fragment-centered database, MSFragDB, by reorganizing the data from the Human Metabolome database (HMDB) and developed an intensity-free searching algorithm to search and rank the most relative metabolite according to the users' input. We also proposed the concept of fragment network, a graph structure that encoded the relationship between the molecule fragments to find close motif that indicated a specific chemical structure. Although based on the same dataset as the HMDB, validation results implied that the MSFragDB had higher hit ratio and furthermore, estimated possible taxonomy that a query spectrum belongs to when the corresponding chemical component was missing in the database. Aid by the Fragment Network, the MSFragDB was also proved to be able to estimate the right structure while the MS/MS spectrum suffers from the precursor-contamination. The strategy proposed is general and can be adopted in existing database. We believe MSFragDB and Fragment Network can improve the performance of structure identification with existing data. The beta version of the database is freely available in www.xrzhanglab.com/msfragdb/.

**Introduction**

Metabolomics that systematically study the cellular chemical compositions and their interaction networks have drawn increasing interests recently for its unique roles in both fundamental biological researches and next-generation precision medicine[1, 2, 3]. To detect the highly diverse chemical components in limited biological samples, untargeted mass spectrometry (MS) is usually applied to acquire the mass-to-charge ratio (m/z) of the molecules in high throughput and sensitivities[4, 5]. Despite advanced experimental techniques are continuing reported to obtain the MS spectra from biological sample with higher coverage[6] or throughput[7], the metabolites identification step still remains time-consuming and challenging and is considered as the major bottleneck to convert the abundant spectra information to biological insights[1, 3, 8, 9, 10]. Compared to the bio-macromolecule such as nucleic acids or proteins that consist of limited number of basic building blocks, the structure of metabolites is highly diverse. To retrieve the structure of the metabolites from the MS spectra, tandem MS (MS/MS) is widely applied in which precursor intact molecule ions are fragmented into parts and the molecular fragments are detected by the secondary MS to form the MS$^2$ spectra. The MS/MS process provides extra structural information of the precursor molecule and served as the basic start point for the following structure retrieve analysis.

Despite large number of algorithms based on machine learning[10, 11] or in-silicon fragmentation[12] has been developed in recent years to estimate the structure of the metabolites from their MS/MS spectra, spectra matching by comparing the experimental data against online database records is still considered as the 'gold standard' for metabolites identification and is also mostly adopted practically. Existing databases such as METLIN[13] and Human Metabolome Database[14] (HMDB) provide invaluable information for metabolites identification but suffer from two major disadvantages: 1) limit component coverage[11]; 2) inefficient spectra similarity matching due to the variations of the MS/MS spectra in different equipment or detection parameters[9, 15, 16], collision energy for example. Global Natural Products Social Networking Library (GNPS), on the other hand, constructs molecular similarity network (MN) to get insights of the structural relationship within the experiment dataset[17, 18]. By doing so, the algorithm focuses on the spectra similarity that acquired by the same researcher with the same equipment and the same batch, thus eliminating the false matching due to the systematic variations. In this way, more chemical structures can be estimated by referring to the well-known node of the MN and prior knowledge about the homologues, even if the component is absent in the database. However, the spectra similarity matching based algorithm intrinsically assuming the secondary MS spectra are originally fragmented from pure precursor component. We argue that, instead, this assumption may fail in analyzing complex biological sample where multiple components with similar *m/z* are fragmented together[19] due to the limited precursor filtering resolution (precursor contamination). To address this issue, mining the relationship information between the fragment ions in a $MS^2$ spectrum may produce extra information about the chemical component identification[20].

In this manuscript, we proposed a fragment-centered method for metabolites identification that considers the fragments of a metabolite component as the building blocks of the molecule and specific fragments can serve as the 'key words' for a series of metabolites. Thus, although the exact chemical component may not exist in the database, its taxonomy still can be estimated by the fragment key words. To prove the concept, we downloaded the dataset of HMDB and reorganized the recorded spectra into a fragment-centered databased named MSFragDB and provided a web-based search GUI in www.xrzhanglab.com/msfragdb/. We randomly selected 40 library spectra from the LipidBlast library and searched them against both the HMDB and the MSFragDB, With the same basic data, the evaluation implied significant matching accuracy boosting by the proposed method. While no correct hit was found in HMDB queries with the database-absent-components, the taxonomy of these components was correctly inferred in MSFragDB with 47% accuracy. We also constructed the molecular fragment network (MFN) to mine the co-existence relationship of the fragments and found the method can provide clues for precursor contamination and estimated the correct components that may co-exist in the same MS/MS spectrum. We re-analyzed the published single-cell MS/MS dataset and found that the $MS^2$ spectra in different cell type can have different profiles, which may due to the precursor contamination. MSFragDB can thus aid the structure estimation in these cases. Overall, the fragment-based database and algorithm provide a new perspective for retrieving structure information from MS/MS spectrum and can be adopted as new searching method in existing databases for match accuracy improvement.

**Results**

**Database construction.** The database was constructed by re-organizing the data from HMDB. For each MS/MS spectrum in the HMDB, fragments info was extracted as individual records. Then the precursor information as well as the reference information including source database and uploader were attached to the corresponding fragment. Finally, the full fragment records were written in MySQL database in the lab Linux service. To provide an easy-to-access interface for data querying, we also developed a website that allowed the researchers to search for possible component taxonomies and structures with their experiment data. Registered users can upload their own dataset to enrich the information of the database. As the uploader meta data will be attached to the fragment record when loading, users can contact the uploader for further information such as the experiment details if necessary. The database will also record the query histories of the registered users for convenient re-visit.

Three major features of the website were designed: single fragment searching, fragment set searching as well as the file-based searching. Single fragment searching (Search One Tab) allows researchers to query all possible fragments within a certain region. This feature may help to estimate the possible components that the fragment belongs to. Fragment set searching (Search Set Tab) comprehensively considering all query results from each fragment of the users' input and estimated and ranked the possible taxonomies and the components, which is comparable to the MS/MS search in other databases. Molecular Fragment Network (MFN) is a default disabled option that introduces the MFN analysis. By enabling MFN, users can investigate the taxonomy similarities among the input fragments to find whether the precursor contamination occurs. File-based searching read the ThermoFisher *.raw file and extracted the fragment information according to the user-input-parameters. The website can be found in www.xrzhanglab.com/msfragdb/.

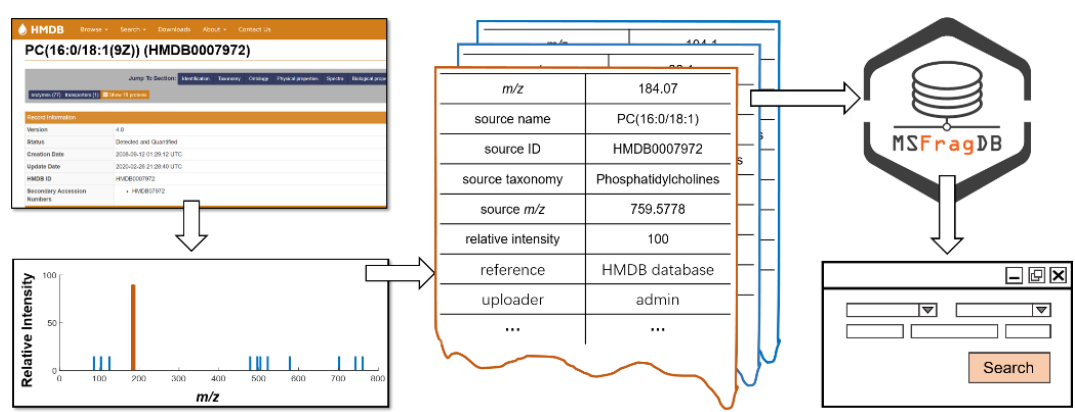

**Fig. 1 The database construction process.**

**Fragment-based searching and ranking.** The search set module is the major feature provided by our online search tool. For each fragment input, various number of hit records can be found in the database and ranking the results to let the most possible candidates appear at the top of the result list is one of the key issues. While the traditional spectrum matching algorithms based on similarity measurements give continuous outcome, the score based on individual fragments can be discrete. Here, we introduced two mechanisms to rank the search results. The first mechanism is a naïve consideration to achieve 'all-hit-priority', which means taxonomies and components that contain the most fragments in query set should be presented firstly. Considering $H_{ijk}$ as a marker for whether input m/z $i$ hit the library component $j$ that belongs to taxonomy $k$. The taxonomy can be scored as

$$T_k^{naive} = max\{C_{1k}, C_{2k}, \dots\} \qquad (1)$$

$$C_{jk} = \frac{\sum_i H_{ijk}}{N} \qquad (2)$$

$$H_{ijk} = \begin{cases} 1 & hit\ record \\ 0 & otherwise \end{cases} \qquad (3)$$

Where N is the number of the query fragments. The second strategy is adopted from the web page searching named TF-IDF scoring. While the web page searching involves topic-webpage-keyword three level organization, the same structure can be found in metabolites identification as the taxonomy-component-fragment organization. Thus, query chemical components by fragments can be considered as the similar process that query webpages by keywords. The

TF term in this case normalized the hit count of fragment $i$ on taxonomy $k$ by the total number of records in the database that belongs to taxonomy $k$. The IDF term, on the other hand, measures the specificity of the fragment. In other words, fragments that generally exist in many taxonomies shows less prediction strength than those only appear in small number of taxonomies. Given $U$ is the set of database records and $U_k$ is the subset that records belong to taxonomy $k$. $S_i$ is the subset that all records that fragment $i$ hits.

$$FT_{ik} = \frac{\sum_j H_{ijk}}{|U_k|} \tag{4}$$

$$IDF_i = log_{10}(\frac{\|U\|}{\|S_i\|}) \tag{5}$$

$$T_k^{FT-IDF} = \sum_i FT_{ik} \cdot IDF_i \tag{6}$$

Where $|\cdot|$ is the size of the set and $\|\cdot\|$ is the number of unique taxonomy in the set. The candidate taxonomies are sorted by the $T_k^{naive}$ and $T_k^{FT-IDF}$. For each taxonomy, the candidate molecules are sorted by the number of hit fragment. Finally, the user interface will present the search results in a three-column hierarchical format to show the information about the potential hit taxonomy, metabolites and hit fragments, respectively (Fig. 2).

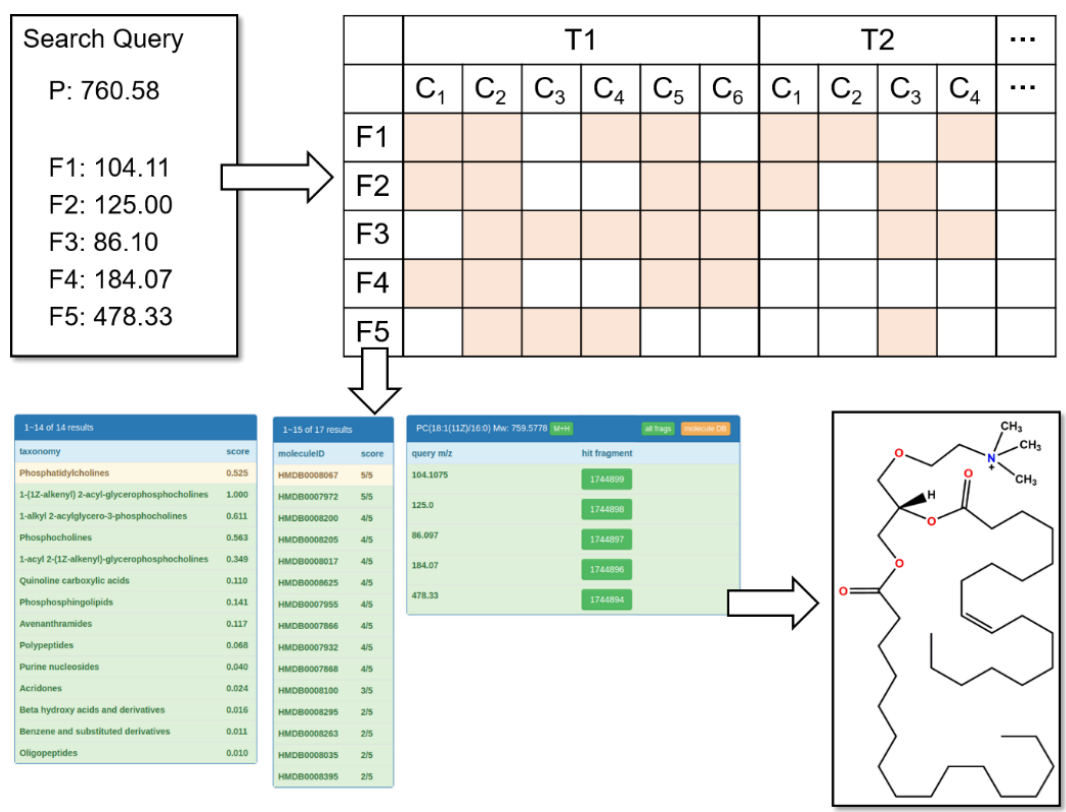

**Fig. 2 Illustration of the searching process.**

**Database evaluation.** We firstly evaluated the database by searching the spectrum within the MSFragDB and 100% of them were correctly annotated by the top item of the resulting taxonomy and component list, indicating that there was no systematic error or bias. Then, we tested the searching accuracy of the MSFragDB by comparing its performance with that of the HMDB, given that they share the same basic data but have different searching strategies.

We randomly selected 40 spectra in the predicted dataset of the LipidBlast library[12] as the lipidomics is one of the most frequently searched areas. Surprisingly, we found none of these spectra was correctly annotated in HMDB, which may due to the large differences of the intensity profile between the query spectra and the library spectra (Fig. 3b). This result indicated that the intensity of the fragment was highly diverse and unreliable for spectrum matching, which was consistent with previous publications[9, 15, 16]. Fragment-based searching strategy, which totally ignores the intensity information, showed significant improvement in correct hit ratio. Specifically, for component that is present in the HMDB (marked as 1 in the GT column in the Fig 3a), top component record achieved 36% accuracy while top5 records achieved 52% accuracy. 72% of the taxonomy can be correctly annotated by the top record while the ratio increased to 88% for top5 records. For component that is absent in the database, the correct hit ratio should be 0% as expected while the taxonomy still can be correctly inferred at 47% accuracy by the top record and 67% by the top5 records. These results show that the fragment-based searching strategy outperformed the traditional spectrum matching algorithm at least in this test case. Also, the strategy adopted in MSFragDB also provides reasonable estimation of the taxonomy of the absent component, which can be helpful in scientific research.

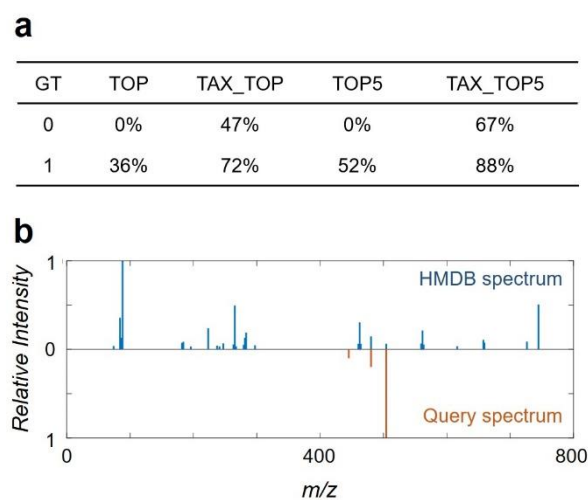

**Fig. 3 Evaluation of the database.** (a) The search accuracy in MSFragDB. GT=0 means the target component is not in the database. (b) A typical query spectrum and its corresponding database spectrum in HMDB.

**Interpret precursor contaminated spectrum.** Precursor contamination can occur in small volume complex sample analysis such as single cell plasma, in which large number of metabolites components exist and separation methods are hard to apply. Traditional methods assume that the MS/MS spectrum is originally from a unique component. However, this assumption may not hold in the complex small sample analysis, where two or more components with similar *m/z* can be fragmented at the same time and recorded in the same spectrum. Here, we took a simple prove-of-concept example in lipidomics analysis. PC (16:0/18:0) (*m/z* = 761.59) and PS (16:0/18:1) (*m/z* = 761.52) have relatively close *m/z* value and can be easily co-fragmented in MS/MS analysis. The resulting spectrum had a characteristic peak with high intensity at *m/z* = 184.07 (Fig. 4a), which made it easily to be considered as a phosphatidylcholine (PC). However, by observing the taxonomy hit matrix $M_{ik} = \sum_j H_{ijk}$ of the query fragment set (Fig. 4b), the hit fingerprint (rows of the matrix) of each fragment indicated roughly two distinct groups existed. The phenomenon can be further conformed by constructing molecular fragment network (MFN), which can be mathematically represented as $G(V, E)$. Each node (V) in the network stands for a fragment and the edges (E) characterize the relationship between two nodes by their weights (Fig. 4c).

$$F_{ik} = \begin{cases} 1 & M_{ik} > 0 \\ 0 & M_{ik} = 0 \end{cases} \quad (7)$$

$$W_{ij} = \begin{cases} \dfrac{\sum_k (F_{ik} \cdot F_{jk})}{\sqrt{\sum_k F_{ik}} \sqrt{\sum_k F_{jk}}} & \sum_k F_{ik} > 0 \ \text{and} \ \sum_k F_{jk} > 0 \\ 0 & otherwise \end{cases} \quad (8)$$

The close related fragment clusters can thus be detected by thresholding the edge weight and social network community detection algorithm[21, 22] (Fig. 4d). Finally, the fragment subset identified by this process can be searched separately in MSFragDB to reveal the true composition of the MS/MS spectrum. This strategy based on MFN and community detection provides a novel way to find clues of the precursor contamination and uncover the compositions of the mixture.

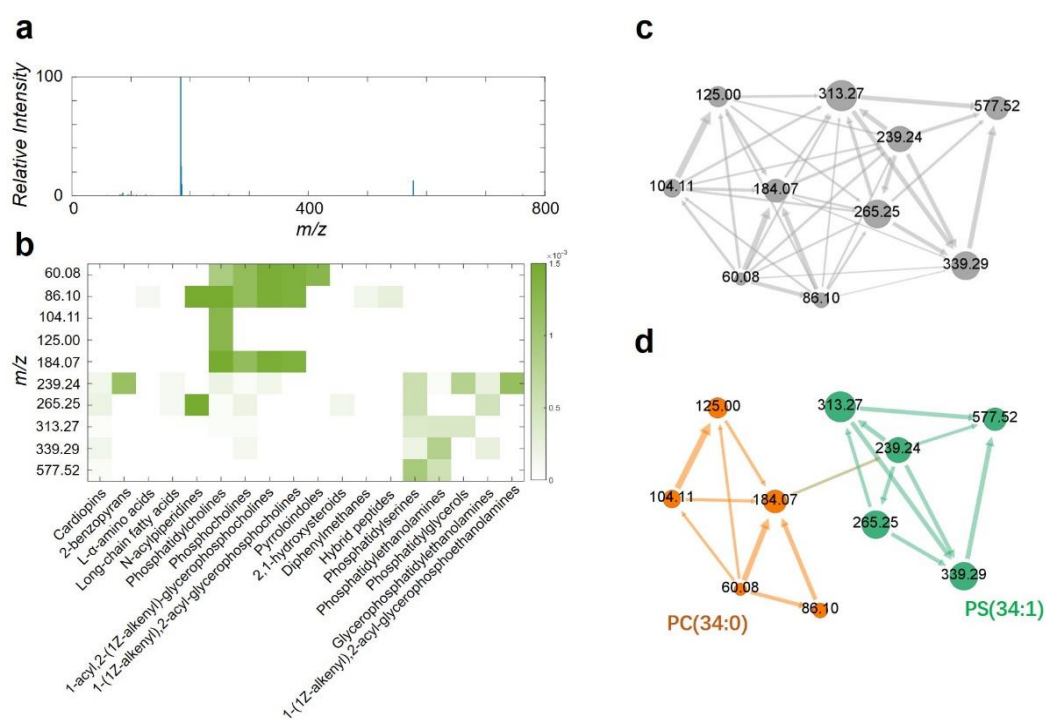

**Fig. 4 Identify precursor contaminations.** (a) The spectrum that co-derives from Phosphatidylcholine (PC) and Phosphatidylserines (PS). (b) Taxonomy hit matrix of the query fragment set. (c) Raw MFN constructed from the hit matrix. (d) Two communities were found by edge weight thresholding and community detection.

**Single cell plasma analysis.** We finally re-analyzed the MS/MS spectra acquired by Pico-ESI-MS where metabolites of both human astrocyte cells and glioblastoma cells were extracted and electro-sprayed into the mass spectrometer. As electrospray time were largely extended, $MS^2$ spectra can be detected for each $MS^1$ peak in a data-dependent manner, and thus abundant MS/MS information can be acquired [23]. However, previous analysis still focused on $MS^1$ spectra such as differentiating cell types according to the $MS^1$ spectra. Here, we focused on the $MS^2$ information of each single cell (Fig. 5a). Interestingly, the $MS^2$ spectra shows significant differences between these two types of cells even after normalizing the sum of the fragment intensity in each spectra to 1 (Fig. 5b). As the normalization step eliminated the abundance information of the metabolites in $MS^1$, if only one chemical component was fragmented in a given precursor channel, there should no differences between these two cell types. Thus, Fig. 5b indicates that multiple components co-fragmentation in a given MS$^2$ spectra may be commonly occurred in single cell metabonomics analysis. These components can have different abundance in different type of cells, leading to different MS$^2$ profiles. In this case, tradition spectrum matching based search algorithms may be misleading as their single-original-component assumption is broken.

However, fragment-centered strategy may be helpful here by mining the relationship among the fragments and uncover their co-existence communities. To find an example case, we filtered the fragment peaks by three criterions: total occurrence times > 5, fold change > 2 and p value < 0.01 (ANOVA) and the filtered peaks were shown in Fig. 5c. We then demonstrated whether the variations of the MS$^2$ spectra with the same precursor $m/z$ were randomly or cell-type-related. Molecular Network analysis was performed (Fig. 5d, precursor $m/z$ = 522.14). Spectrum similarity was measured as described before[24] with 5 strongest peaks selected, 0.01 Da alignment threshold and edge weight lower than 0.95 were eliminated. The resulting MN was plotted with force layout[25] which shows that the two type cells clustered separately. This result indicated that the MS$^2$ spectra of the single cell samples had higher similarity within the cell types than that between different types. Thus, the variations were cell-type-related, which confirm the assumption we made above. Two typical spectra were plotted in Fig. 5e. The spectrum corresponding to human normal astrocyte cells (blue color) showed clear fragmentation pattern that easy to be interpreted: all peaks such as 86.10, 125.00 and 184.07 were related phosphatidylcholine structure. However, the spectrum corresponding to cancer glioblastoma cells showed much complex pattern while still preserving phosphatidylcholine-related peaks. We selected 10 strongest peaks and searched them in MSFragDB. Molecular Fragment Network analysis showed that 3 fragment clustered were formed and the top-hit records for each cluster were LysoPC, Flavonoid-O-glycosides and Furospirostanes, respectively. These results imply that the precursor-contamination may frequently occur in the MS analysis of complex small volume samples and MSFragDB can help to identify the right component composition with a fragment-centered strategy.

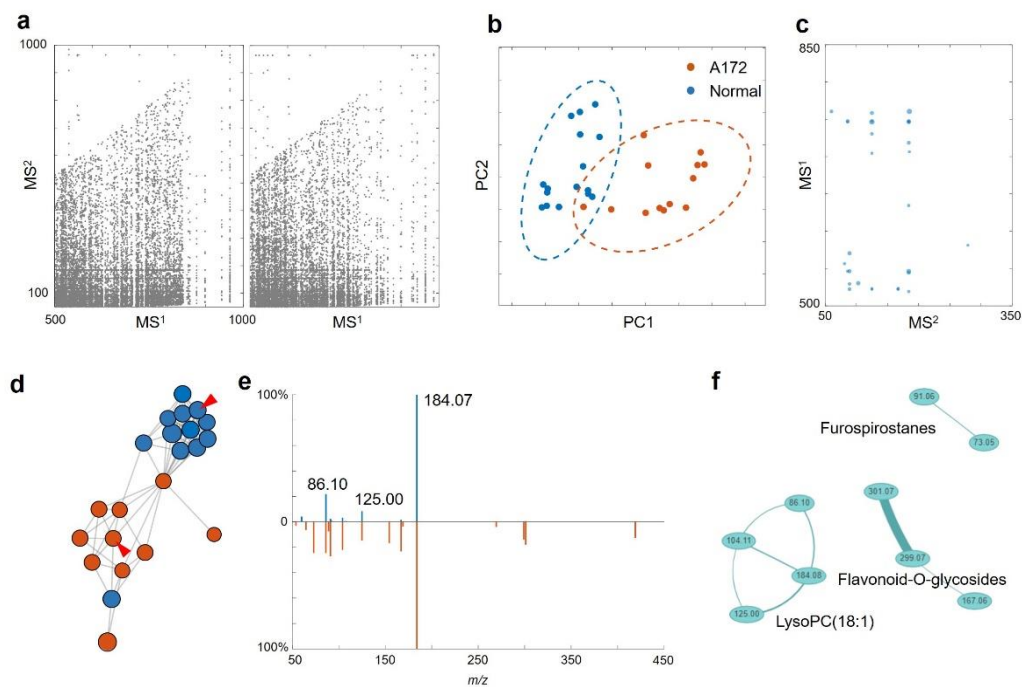

**Fig. 5 Identify precursor contaminations.** (a) Typical MS/MS map of A172 (left) cells and Normal (right) cells. (b) PCA analysis of the single cells. (c) Map of significant different MS/MS peaks (fold change > 2, -lg(p) > 2). (d) Molecular Network of the precursor with m/z=522.14. The orange color and blue color indicate A172 cells and Normal cells, respectively. (e) MS/MS spectra correspond to the nodes

indicated by red arrows in (d). (f) Fragment network constructed by MSFragDB and the suggested metabolites.

**Discussions**

Here, we proposed a fragment-based strategy to aid the metabolites identification. By constructing a fragment-centered database named MSFragDB and evaluating its performance, we proved the hit accuracy improvement of the fragment-based strategy, which may encourage existing databases to adopt it to improve the matching performance. As the intensity of the fragment peaks in MS$^2$ spectra has large variations in different equipment settings, the fragment-centered matching strategy may be more reliable than the spectrum-level similarity measurement.

While large number of MS spectra can be acquired in short time by advanced equipment and analysis methods, the MS/MS spectra dataset collected can carry more and more abundant information. The MFN analysis proposed here can be an efficient method for the large dataset mining and fragment relationship extraction. It's reasonable to assume that the fragments of molecules have certain relationships. For example, fragments of the characteristic structures of a chemical taxonomy may more likely coexist in MS/MS spectra. Mining these relationships can simplify the MS$^2$ profile and aid the following spectra interpretation and structure identification. Moreover, our analysis on the single cell MS/MS spectra dataset suggests that the precursor contamination can frequently occur in the fragmentation processes of the small volume complex biological samples, which breaks the basic assumption of the spectrum-level matching algorithms. The fragment-centered strategy is essential for uncovering the right metabolites.

We also build a community prototype in the website by allowing users upload their own experiment data and leave their information in the personal page. By doing so, researchers can contact with each other for further cooperation. However, current records in MSFragDB are originally from the data in HMDB, which limits its performance. Adding more data source should improve its searching accuracy. Further works may also include adding batch searching method to meet the demand of high-throughput data analysis.

The manuscript is prepared for preprint submission. The online website proposed here was developed by the first author for prove-of-concept. It may suffer from unexpected bugs or accidents such as unreliable network connection in the lab. Any comments, suggestions or bug reports are welcome and can be submitted via e-mail of the first author (zhaohs16@mails.tsinghua.edu.cn) or the corresponding author (xrzhang@mail.tsinghua.edu.cn; sczhang@mail.tsinghua.edu.cn).